\documentclass{article}
\usepackage{geometry}

\usepackage[utf8]{inputenc}

\usepackage{cite}

\usepackage{nameref,hyperref}

\usepackage[right]{lineno}

\usepackage{microtype}
\DisableLigatures[f]{encoding = *, family = * }

\usepackage{changepage}

\usepackage[aboveskip=1pt,labelfont=bf,labelsep=period,singlelinecheck=off]{caption}

\makeatletter
\renewcommand{\@biblabel}[1]{\quad#1.}
\makeatother

\usepackage{color}

\definecolor{Gray}{gray}{.25}

\usepackage{graphicx}

\usepackage{sidecap}

\usepackage{wrapfig}
\usepackage[pscoord]{eso-pic}
\usepackage[fulladjust]{marginnote}
\reversemarginpar

\usepackage[skip=5pt]{caption} 
\usepackage{subcaption}
\usepackage{float}

\usepackage{tabularx}
\usepackage{booktabs}

\usepackage{authblk}

\usepackage[title]{appendix}

\usepackage{booktabs,siunitx}

\title{Data Commons}

\author[1]{R. V. Guha*}
\author[1]{P. Radhakrishnan}
\author[1]{B. Xu}
\author[1]{W. Sun}
\author[1]{C. Au}
\author[1]{A. Tirumali}
\author[1]{M. J. Amjad}
\author[1]{S. Piekos}
\author[1]{N. Diaz}
\author[1]{J. Chen}
\author[1]{J. Wu}
\author[1]{P. Ramaswami}
\author[1]{J. Manyika}

\affil[1]{Google}

\begin{document}

\maketitle

\begin{abstract}
Publicly available data from open sources (e.g., United States Census Bureau (Census) \cite{censuswebsite}, World Health Organization (WHO) \cite{whowebsite}, Intergovernmental Panel on Climate Change (IPCC) \cite{ipccwebsite}) are vital resources for policy makers, students and researchers across  different disciplines. Combining data from different sources requires the user to reconcile the differences in schemas, formats, assumptions, and more. This data wrangling is time consuming, tedious and needs to be repeated by every user of the data. Our goal with Data Commons (DC) is to help make public data accessible and useful to those who want to understand this data and use it to solve societal challenges and opportunities. We do the data processing and make the processed data widely available via standard schemas and Cloud APIs. Data Commons is a distributed network of sites that publish data in a common schema and interoperate using the Data Commons APIs. Data from different Data Commons can be ‘joined’ easily. The aggregate of these Data Commons can be viewed as a single Knowledge Graph. This Knowledge Graph can then be searched over using Natural Language questions utilizing advances in Large Language Models. This paper  describes the architecture of Data Commons, some of the major deployments and highlights directions for future work.    
\end{abstract}

\section{Introduction}

Data is vital to understanding the world and improving public welfare. Increasingly complex global challenges require scientists and policymakers to synthesize insights and anticipate future scenarios from a diverse collection of data sources.  For example, responding to the climate change crisis requires understanding the second-order effects of temperature changes on food production, shelter, healthcare, and other aspects of life. Importantly, these analyses often cut across the boundaries of existing data collections, because they require information from different domains often collected by siloed government or research entities. Consider, for instance, American public health officers attempting to predict which hospital systems are most likely to be gradually overwhelmed by the effects of increasing temperatures. A starting point would be to identify counties that satisfy all these criteria: (1) have a high prevalence of heat-sensitive health conditions (e.g., heart diseases, old or young aged populations), (2) expected to experience significant temperature increases over the next decades, (3) have frail energy infrastructure which will not withstand a high heat event (e.g., previous blackouts or energy production per population), (4) have weak public health infrastructure (e.g., doctor per person or distance to closest medical center). 

As a practical matter, however, conducting any one of the above analyses can require extraordinary effort. The public health officers must first identify which databases possess the relevant pieces of information (in this case for the United States: IPCC \cite{ipccwebsite}, CDC \cite{cdcwebsite}, Census \cite{censuswebsite}, United States Department of Energy (DoE) \cite{doewebsite}, and more) and which temperature prediction models they want to use (there are over 25 options). They must then download all the datasets, determine how they should be merged, and then perform said merging. Through all of this, they must account for the fact that these databases are available in different data formats, represent important variables in different ways, and capture information at different times or spatial granularities (e.g.,  IPCC data is provided at geospatial grid granularity, while the other datasets include data at county, census tract or city granularity). To say nothing of the fact that they must know (or learn) the computational tools necessary for performing these steps, and repeat it all if they wish to change their analysis (e.g., using a temperature forecast not contained in the data they downloaded). 

The challenges that arise in performing this kind of basic data analysis present several drawbacks. First, they induce extraordinary expenditures of time and effort. Studies have shown that data scientists spend up to 80\% of their time on basic data cleaning, merging, and management tasks \cite{cleaningbigdata}. This is an inefficient use of expert hours and limited research funding. Second, they limit who can perform such analyses, and thus, who can benefit from them. Data analysis is most valuable when conducted by front-line workers closest to the problems --- individuals like local public health officers or managers at a food pantry. Yet, the level of technical sophistication required to perform the necessary data management operations prevents these stakeholders from engaging in data analysis, thereby depriving them, and many communities, of the benefits of data-based decision making or advocacy.  

To address these challenges, we developed Data Commons. The Data Commons approach is to manage the data wrangling once and make the processed data widely available via standard schemas and Cloud APIs in an open source stack. Anyone with internet access should be able to derive the full benefit of all this data.

Data Commons has three major components:
\begin{itemize}
    \item A framework for data publishing, which includes a set of schemas, based on Schema.org \cite{schemaorg}, and a set of APIs. Data made available via these schemas and APIs by one site can easily be combined with data from other sites. Multiple reference implementations of this framework are described below.

    \item A large, public Data Commons (the Google Public Data Commons \cite{dcwebsite}) that aggregates normalized data from a wide range of government and NGO sources, intended to jump start the ecosystem.

    \item A suite of tools that work with any site that publishes data in these schemas and APIs.
\end{itemize}

\paragraph{Organization}
The rest of this work is organized as follows: first, we review the relevant bodies of work in Section \ref{sec:related}. Next, in Section \ref{sec:arch}, we describe the architecture of Data Commons, including schemas, APIs, implementations of Data Commons and tools that work across all Data Commons instances. We describe the layering of data across Data Commons instances in Section \ref{sec:federation}, and finally provide examples of implemented Data Commons instances in Section \ref{sec:examples}.

\section{Related Work}
\label{sec:related}
We highlight three existing types of systems which bear similarities to Data Commons, and discuss the ways in which Data Commons is different from these efforts.

\paragraph{Knowledge Repositories}
Starting with work as early as the 1980s, there have been a number of repositories of processed data. Early work such as Cyc \cite{cyc} was both more ambitious in aiming to capture common sense and less ambitious in the size of the repository. There have been a number of more recent projects that bear greater similarities to Data Commons. One such example is the World Wide Telescope (WWT) \cite{wwtelescope} from Microsoft. In \cite{jimgraywwt1}, Gray et al. describes aggregating astronomical data from many data sources into a single SQL server database and making this available over the Internet. Prior to this, the only way of accessing this data was for a researcher to spend time at the observatory, something which required significant resources. Unlike the WWT, which included data only about astronomy, Data Commons includes data about a much wider range of topics, including demographics, economics, climate, health, crime, and others. 
Wikidata \cite{wikidata} is a crowd sourced open repository of some of the structured data in Wikipedia. Like Data Commons, it uses Knowledge Graphs as the data model. Wikidata is mainly in service of structured data attached to Wikipedia and therefore tends not to maintain detailed timeline data and deep statistics or observations. Data Commons does use Wikidata for place encoding for things like administrative area recognition. 
Google’s Search Knowledge Graph \cite{googlekg} (and similar constructs from Microsoft, Apple, etc) have followed the same approach as the World Wide Telescope, with a broader focus on the kinds of search queries posed to the portals supported by these systems. Wolfram Alpha \cite{wolfram-alpha} is another system that aggregates data and provides a natural language interface. Like Google's Search Knowledge Graph, it is a single closed and proprietary database. Google’s Search Knowledge Graph tends to be more pointed towards consumer use cases and thus lacking in detailed statistical data. 

  Data Commons differs from these in two important respects:
  \begin{enumerate}
      \item  Unlike these systems, which are all single monolithic databases, Data Commons is not a single database instance, but rather, an ecosystem of distributed and interoperable DC instances that are built to be open.
\item The Public Data Commons built by Google, available at DataCommons.org, is primarily focused on public data that is required for solving societal challenges. This instance is meant to store a large number of available datasets to help illustrate the value of interoperable data and encourage more data providers to create their own Data Commons instances. 
 \end{enumerate}

\paragraph{Linked Data} Linked Data promotes the idea of a distributed set of knowledge graphs \cite{bernerslee1, bernerslee2}. Data Commons differs from these efforts in several ways. First, Linked Data encourages different sources to use the same identifiers for entities (nodes in the graph), but allows different sources to use different schemas, i.e., different set of types, different attributes or arc labels, etc. Thus, two sources may use different terms to refer to the same properties (e.g., population), and may even use different formalisms for key constructs such as time and provenance. This makes it substantially more difficult to combine data from these different sources. In contrast, following Schema.org \cite{schemaorg}, Data Commons provides a common set of types and properties (among other aspects). Second, Linked Data requires that all sites use the same identifiers for all entities. Data Commons, in contrast, allows for data sources to use different identifiers, and relies on Reference by Description \cite{rbd} to resolve entity references. Third, Linked Data does not promote common vocabularies for the representation of time and provenance. Data Commons has created these vocabularies and promotes their usage.

\textbf{Dataset repositories}: With the advent of cloud computing, major cloud vendors have started providing repositories of datasets. Examples include Amazon’s Data Exchange \cite{awsdataexchange}, Azure Open Datasets \cite{awsdataexchange} and Google Cloud’s Datasets \cite{gclouddatasets}. Similar open efforts include Dataverse \cite{king1, king2} and data.gov \cite{datagov}.

The principal aim of Dataverse and data.gov is to enable users to efficiently search for relevant datasets. The cloud data repositories go further and eliminate the need to download the data (in the case where that cloud is used for the analysis). However, none of these systems specify how the data is to be formatted, enforce any standardization, or alter how users actually engage with data. Thus, users must still perform traditional forms of data merging, cleaning, and standardization. In Data Commons, however, such steps have already been taken.

\section{Data Commons Architecture}
\label{sec:arch}
In this section, we describe the Data Commons architecture. This architecture can be analogized to the Web. On the Web, users can find different websites which vary on dimensions like topic, accessibility, and terms-of-service. Despite these differences, all websites make content available under a small number of common formats (e.g., HTML, CSS, JavaScript, and others). This enables the same set of tools---like web browsers from different companies, web servers from multiple companies, or applications like Google Maps---to be used across them. These websites also all use the same set of protocols for accessing the data (HTTP) and a uniform addressing scheme (URLs) which makes it possible for a page on one site to link to a page on another site. 

The Data Commons ecosystem follows a similar structure. Each Data Commons is comparable to a website, and may have a specific topical focus (e.g., food security like Feeding America \cite{feedingamericawebsite}), a specific level of accessibility (e.g., researcher access only), and unique terms-of-service. Just as websites operate on common formatting and communication protocols, all Data Commons’ operate on a common set of schemas. While reliance on HTML and HTTP allows users to seamlessly view and navigate between different websites, reliance on the Schema.org \cite{schemaorg} plus Data Commons schemas allows users to rely on a single API for downloading and accessing different data about common entities. 

The imposition of common data access APIs and representation schemas provides several benefits. First, different Data Commons can be implemented in radically different ways, yet still seamlessly interface with each other. For instance, existing Data Commons implementations range from ultra lightweight in-browser JavaScript implementations, to mid-range in-memory database based implementations, to super scalable BigQuery \cite{bqwebsite} implementations. Implementation decisions can thus be made on the basis of infrastructure availability, without concern of diminishing data usability. Second, reliance on common schemas and APIs means that data tools, scripts, or applications developed for one Data Commons work on all Data Commons. This includes, for instance, tools for visualization, analytics, machine learning, and natural language (e.g. search or “chat”) interfaces. 

Most importantly, however, the common schemas and APIs mean it is possible to easily join data across these different instances of Data Commons. We next discuss this in more detail.

\subsection{Schemas}

The core data model underlying DC is a Knowledge Graph (KG). This KG is built upon the data model used in Schema.org \cite{schemaorg}, a standard that is used by over 50 million sites. Schema.org---an extension of Resource Description Framework (RDF) Schema \cite{rdf}–starts with a directed labeled graph model and adds the following:
\begin{itemize}
    \item A type system where the types (or classes) are themselves nodes in the graph.
    \item Arc labels (or properties) which are also nodes in the graph. Properties have one or more types as domains and ranges.
\end{itemize}

To this, we add uniform representations for the following:

\begin{itemize}
    \item  Different ways of specifying spatial regions, including S2 cells, arbitrary polygons, and others.
\item Ontologies for the kinds of entities commonly referred to in national statistics, health, economics and related datasets. This includes:
\begin{itemize}
    \item Administrative areas, such as countries and their breakdowns (e.g., cities, states, provinces, etc). 
    \item Attributes commonly covered in these datasets, including attributes of people (demographics, employment, health conditions, etc.).
\end{itemize}
\item Temporal aspects of statements (e.g., the population of Croatia in 2019 was 4.08 million).
\item A uniform mechanism for associating provenance with statements at a very fine granularity.
\item Simple composition mechanisms: Traditional knowledge graphs, which are a small subset of classical first order logic, lack composition mechanisms. We add some basic composition mechanisms which allow us to have a uniform, scalable representation for quantities (as a composition of a unit of measure together with a value), for intervals (of time, etc.), locations (e.g., latitude and longitude), etc.
\item A mechanism for representing the decomposition of composite variables into component dimensions (e.g., “population of hispanic women” $\rightarrow$ count of entities of type=Person, gender=Female, race=Hispanic).
\end{itemize}

The above representation formalisms form the ‘core.’ Given the open ended scope of Data Commons, together with its distributed nature, it is inevitable that there will be multiple independent, overlapping, topic specific vocabulary extensions to the core. However, since the suite of tools depends on adherence to the core vocabulary, we expect conformance to the core vocabulary.

\subsection{APIs}
This section describes the Data Commons API. Data Commons instances must provide a basic graph navigation API. The decision to adopt a simple API was motivated by observations about the tradeoffs of existing types of APIs. There are many graph databases on the market (e.g., Neo4j \cite{neo4j}, Amazon Neptune \cite{awsneptune}), which offer a wide spectrum of APIs (ranging from very expressive query languages such as SPARQL \cite{sparql}) to more restricted APIs such as GraphQL \cite{graphql}. However, experience with the Web shows that while expressive query languages such as SQL are useful in highly controlled environments, for wide public usage, simpler, easier to understand, more predictive APIs (such as the extremely simple one in HTTP) are more successful.

The Data Commons API is a simple graph API which enables a program to navigate the Data Commons graph with a combination of calls, together with some utility functions.
\begin{itemize}
\item Given a node, return the labels of all the arcs in / out from that node
\item Given a node and a label, return the target / source of all the triples with that label and that node as the source / target.
\end{itemize}

Both these calls can be augmented with parameters to restrict the answer by provenance. 

Each entity in Data Commons has an associated `dcid' (Data Commons Identifier), which is used to refer to it in the API calls for navigating the immediate neighbourhood of the graph around it. An important step for a Data Commons user is identifying the dcid's of the entities they care about. Since there can be millions of entities, it is unreasonable to expect dcid's to be explicitly expressed in code.

We solve this through reference by description. Data Commons provides an API which maps simple descriptions of entities (e.g., the country of Georgia, as opposed to Georgia, USA) to dcid's. Currently, this API only supports a limited range of descriptions, in a structured form, e.g.,
\[
    \verb|{ name: ``Georgia''; typeOf: Country }| \rightarrow \verb|{ dcid: country/GEO }|
\]
In the future, we expect to support more open ended descriptions stated in natural language (e.g., ``the country Georgia'').

While the core graph model (and hence the graph API) is general enough from a representational expressiveness perspective, we have found it very convenient to have APIs based on higher level abstractions such as time series.

\subsection{Implementations}
   Data Commons is defined not by a particular implementation, but by a set of schemas and protocols. Depending on the scale of data and usage, different implementations make more sense. We briefly describe three implementations that illustrate the range of options. They differ primarily in the amount of data they can store. 

\begin{itemize}
\item A very lightweight JavaScript based implementation that runs in the browser. We have found this useful for creating a transient DC view of some data that might be available as a CSV or some other simple format. Note that even in this very lightweight implementation, thanks to the layering, all of the data in the base layers is available
\item In the last few years, simple relational database systems such as SQL Server \cite{sqlserver}, Oracle Database \cite{oracledb}, PostgreSQL \cite{postgresql} etc have shown the ability to store many hundreds of gigabytes of data. We have built an open source DC implementation that relies on this as the persistent data storage layer. 
\item An implementation that uses a combination of BigQuery \cite{bqwebsite} and Bigtable Cache \cite{btwebsite} to provide the backend for a Data Commons that can store Knowledge Graphs with trillions of triples. BigQuery is used to store the data and Bigtable cache provides a caching layer for fast serving of frequent queries.

\end{itemize}

\subsection{Tools}
 We have built a number of tools that can work across all DCs, i.e. they only assume that DC provides the basic APIs and the core vocabulary. These tools include:

\begin{itemize}
\item Visualization tools: We have developed a suite of visualization tools, including map based, time series based, scatter plot based visualizations.
\item A natural language tool that allows a user to explore a Data Commons using natural language queries.
\item In addition to the basic APIs, a suite of tools for easy access of the data into Google Sheets, as CSVs, libraries for access from Python notebooks, etc.
\item A suite of tools for helping normalize data for addition into a Data Commons, including a simple user interface to import a single CSV file, an instance of Data Commons that runs in the browser to visualize the prepared data, and open-sourced Python libraries to aid data processing.
\end{itemize}

\begin{figure}[H]
    \centering
    \begin{subfigure}[b]{0.3\textwidth}
        \frame{
            \includegraphics[width=\textwidth]{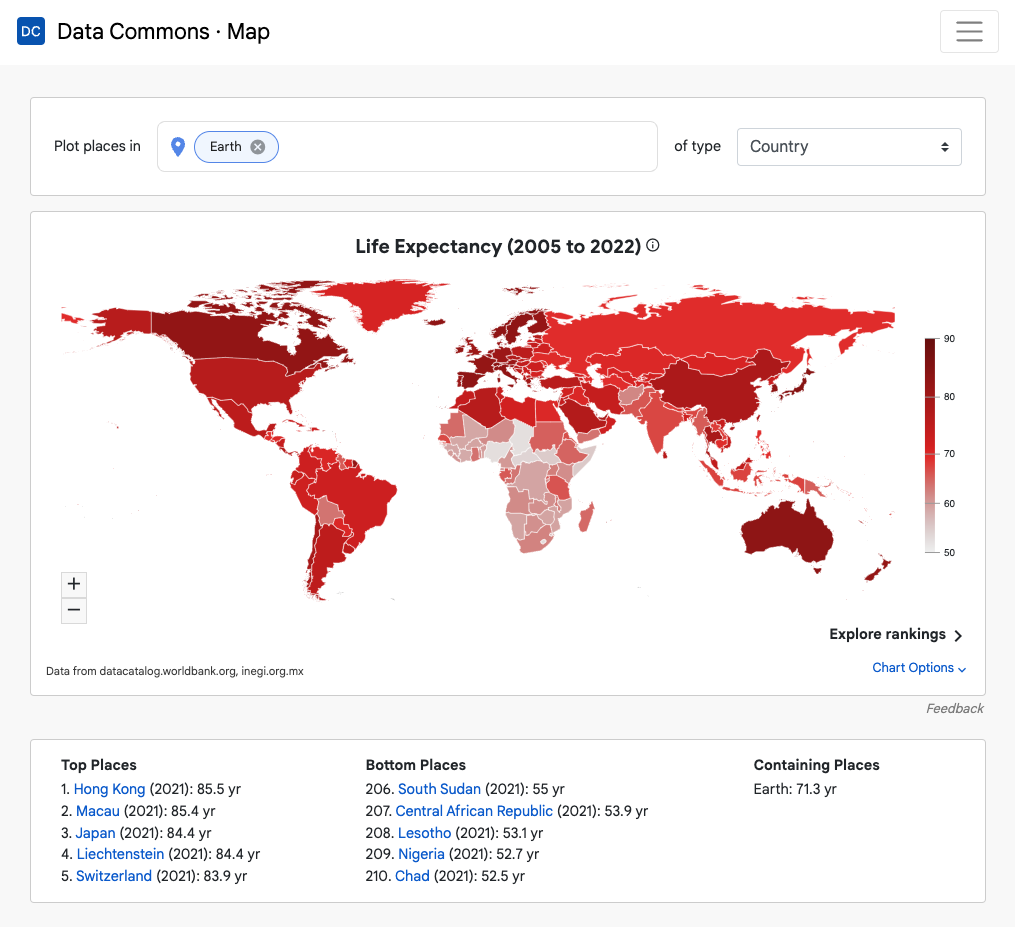}
        }
        \caption{Map Explorer}
        \label{fig:dc-map}
    \end{subfigure}
    \hfill
    \begin{subfigure}[b]{0.3\textwidth}
        \frame{
            \includegraphics[width=\textwidth]{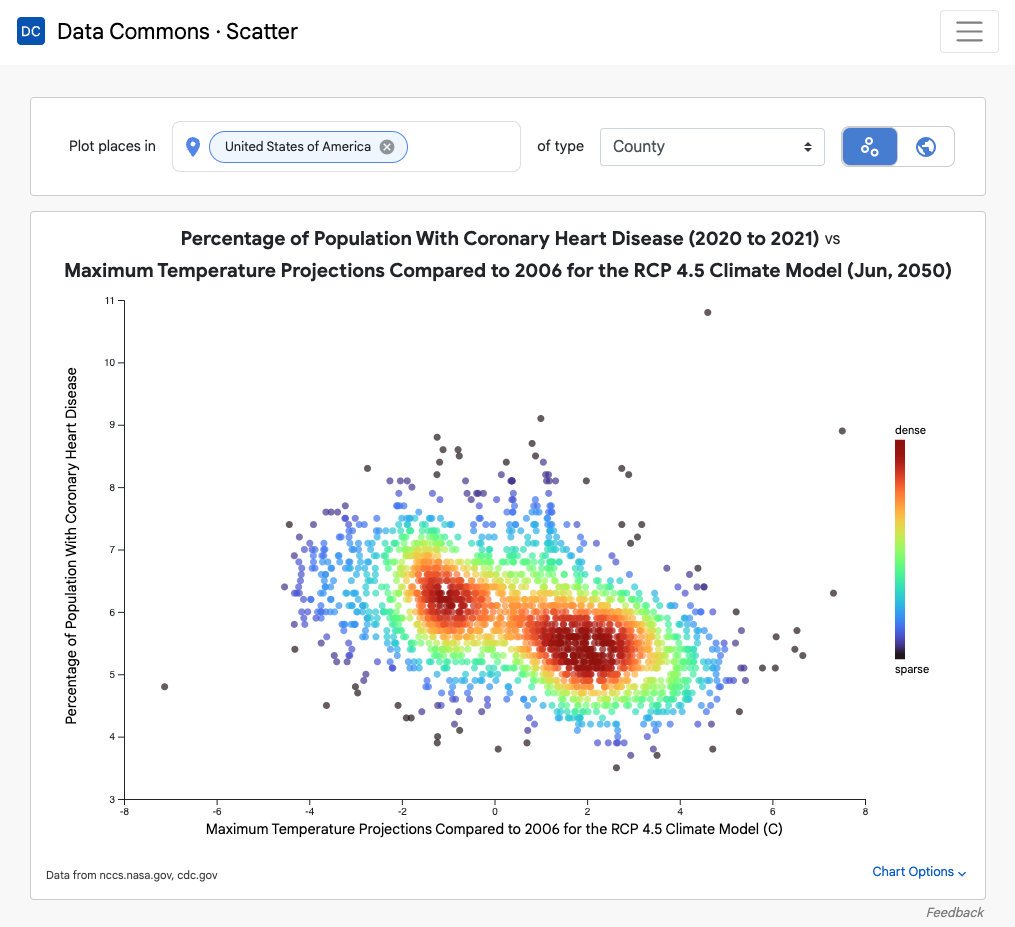}
        }
        \caption{Scatter Plot Explorer}
        \label{fig:dc-scatter}
    \end{subfigure}
    \hfill
    \begin{subfigure}[b]{0.3\textwidth}
        \frame{
            \includegraphics[width=\textwidth]{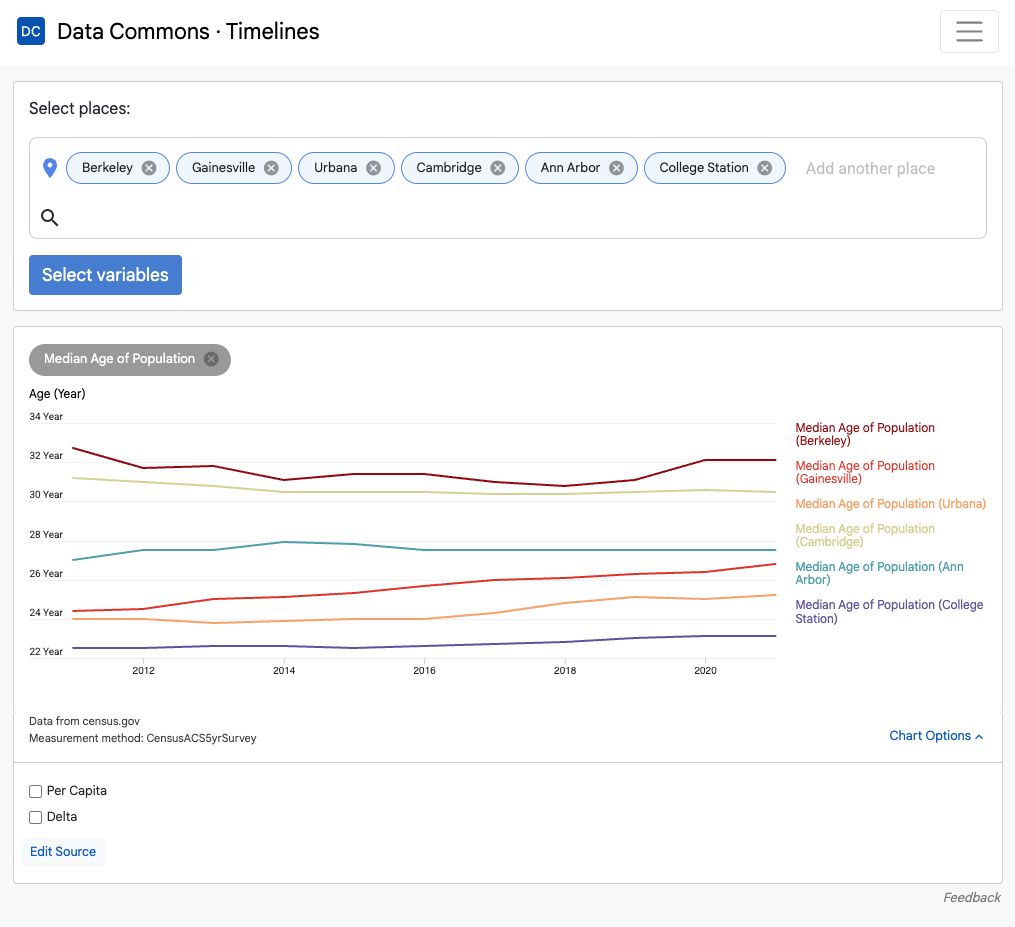}
        }
        \caption{Timelines Explorer}
        \label{fig:dc-timeline}
    \end{subfigure}
    \caption{Visualization tools in Data Commons.}
    \label{fig:dc-viz}
\end{figure}

\section{Combining Data Across Data Commons}
\label{sec:federation}
There are two main incentives for someone wishing to publish data to make it available in a Data Commons:

\begin{itemize}
\item There is a growing suite of visualization and analysis tools that are available for data made available via Data Commons schemas and APIs.
\item Data from different Data Commons can easily be combined / joined without further data wrangling, increasing the value of all the data sets.
\end{itemize}

Elaborating on the second point, one of the key incentives for making data available in a Data Commons is that it becomes much easier to join the new data with data from other Data Commons. A user can make separate API calls to the different Data Commons and perform the join easily after retrieving the Data. The DC architecture provides an easier alternative for a Data Commons to offer a view of its data that is ‘pre-joined’ with data from other Data Commons. This is done by a Data Commons having one or more base Data Commons that it recursively makes the API calls to, merging the results from them with its own results. This kind of ‘layering’ makes it very easy for someone with data to provide a view of that data integrated with a much larger corpus of open data. Layering is an important part of the Data Commons design.

\begin{figure}[H]
    \centering
    \includegraphics[width=0.75\textwidth]{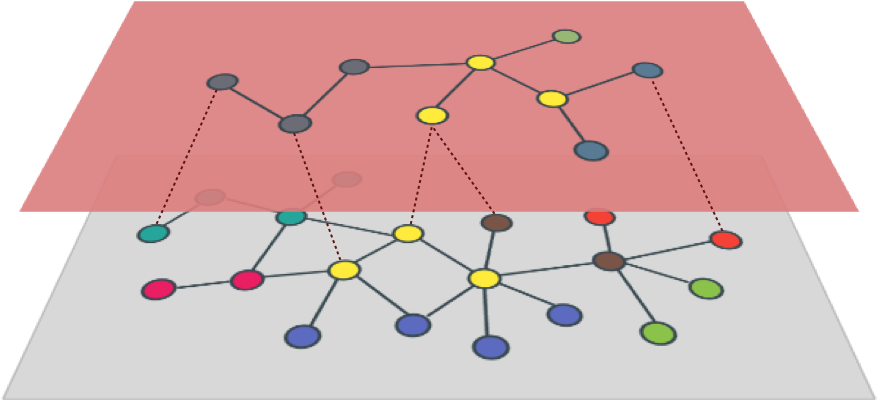}
    \caption{An illustration of the Feeding American Data Commons overlayed on the Google Public Data Commons}
    \label{privatedc}
\end{figure}

The following example illustrates the kind of analysis  that is vastly simplified by Data Commons and the challenge in getting there. Feeding America \cite{feedingamericawebsite} is a charitable organization that is involved in the operation of more than half the food pantries in the United States. Over the years, they have developed a set of indices for measuring food insecurity --- the Meal Gap Index \cite{mapthemealgap}. They estimate values for these indices for every county in the US. They would like to understand the correlation between this index and other data such as income, education, employment, insurance, and more. This other data comes from a large number of different sources. 

The promise of Data Commons is that if Feeding America makes its Meal Gap Index data available via a Data Commons and the other data is available in a Data Commons, then this analysis should be very easy. There have been many proposals in the past with similar outlines, i.e., if everyone makes their data available in easily interoperable schemas and formats, a large class of tasks will be greatly simplified. However, the problem with almost all of these proposals is that the ‘other’ data is not yet available in a Data Commons and until that is available, it might not be worth the effort for someone like Feeding America to make their data available as a Data Commons. In other words, we have to somehow jumpstart the ecosystem.

To help with this jumpstarting, we have created a large Data Commons (available at DataCommons.org \cite{dcwebsite}) which includes a wide range of data about demographics, economics, health, bio-medicine, climate, crime and other topics from over 120 different national, state and inter-governmental agencies. We describe this in the next section.

\section{Examples of Data Commons}
\label{sec:examples}
 In this section, we briefly describe a few Data Commons, starting with Google Public Data Commons \cite{dcwebsite}. 

\subsection{Google Public Data Commons}

 The Data Commons that is available at the site datacommons.org --- Google’s Public Data Commons (GPDC)--- is a large knowledge graph of public data that is targeted at helping solve some of society's biggest challenges, including adapting to climate change, battling different forms of inequity, and understanding the economic trajectory of societies over the last decades. Consequently, some of the topics covered include:

\begin{itemize}
\item {\bf Economics:} We need a deep understanding of the economic context of communities at different granularities, from the country level all the way down to the census tract level, in order to better understand the different kinds of challenges and opportunities facing them. Consequently, a significant fraction of GPDC is economic data, ranging from employment, income, output of different sectors, poverty, trade, GDP, and more. Sources for some of these economic datasets include the World Bank, country census departments, the US Bureau of Labor Statistics, Eurostat, and more. A user can ask Data Commons about Economics with queries such as [Show me the breakdown of businesses by type in the US] or [Tell me about the economy in Brazil].

\item {\bf Demographics:}  Demographic breakdowns around age, gender, income, race, religion, and more are another essential area that can help policymakers, journalists, and researchers understand how different populations in a society are benefiting or suffering. GPDC includes a wide range of demographic data from a large number of different country level and international sources, mostly from national census bureaus, but many datasets about economic outcomes, health outcomes, political representation, and more are all broken down by demographics such as age, gender, income, race, and religion. Sample questions about Demographics include [Which counties in the US have the most immigrants] and [What are the demographics of divorced people in California].

\item {\bf Public health:} Public health decisions rely on large, longitudinal datasets to help identify the value of different interventions. Today, GPDC includes public health data from a small number of sources such as the country Health Ministries such as the US Center for Disease Control, the World Health Organization, the World Bank, the Organisation for Economic Co-operation and Development, and more. This data allows us to understand condition prevalence, budgeting around health, insurance status, or immunization status of different populations as a few examples. For our health data, we are actively looking to expand coverage. One can ask Data Commons about health such as [How does life expectancy vary across countries in Africa] or [Which counties in Illinois have seen the greatest increase in the number of cancer deaths]. 

\item {\bf Climate:} Climate change is one of the biggest challenges of our times. In addition to data about Greenhouse Gas emissions, air quality, climate related disasters, pollution, and more, Google’s Public Data Commons has curated a unique climate projection model. Climate projection data is largely open and available from the IPCC. Yet, it contains data at different spatial resolutions, from modeling organizations spread across the world. Answering a simple sounding question like ‘how hot might it be, with a given likelihood, in a certain period of time, in a certain place’ requires a level of computational resources unavailable to most. We have done this processing work and incorporated it into the GPDC as an open source output for the world. You can ask Data Commons about different Sustainability topics such as [What are the projected temperature extremes across the counties in Montana] or [Which countries have the most greenhouse gas emissions]. 

\item Data Commons contains many more topics of data around housing, education, crime, commutes, equity, and more. Data Commons is today able to answer questions such as 
[Which states have the most college educated in the US], [How have house rents in Mountain View changed] or [Tell me about maternal mortality across countries in the world].

The real value of Data Common’s singular knowledge graph comes to fruition in the ease with which one can compare disparate, historically siloed datasets. For example, many public health researchers want to compare educational outcomes, income, crime and safety, and other “social determinants of health” with condition prevalence for specific demographic communities. While historically this would require a great amount of money and resources spent in hiring computer scientists and data scientists to clean, forage, and analyze this data, Data Commons has made this as simple as asking the kind of questions shown in table 1.

\end{itemize}

\begin{table}[]
    \centering
    \begin{tabular}{l}
    \toprule
        \textbf{Some of the natural language queries that can be answered} \\ \midrule
         ``Compare mortality in the US vs. Spain vs. Nigeria''   \\ 
         ``What is the correlation between obesity and unemployment in the US counties''  \\ 
         ``How does child immunizations correlate with life expectancy in Asia''  \\ 
         ``Compare health conditions vs. median age in Alameda County''  \\ 
         ``Does obesity correlate with lack of sleep in US counties''  \\ 
         ``Which countries have the best immunization rates in the world''  \\ 
         ``Tell me about infant mortality in the world''  \\ 
         ``How does life expectancy vary across US states''  \\ 
         ``Tell me about maternal mortality across countries in the world'' \\ 
         \bottomrule
    \end{tabular}
    \caption{Sample of natural language queries, answerable by Google Public Data Commons, which illustrate the ability to easily combine data across different topics.}
    \label{tab:my_label}
\end{table}

It is important to note that all of the data in GPDC comes from public sources already available on the web with Creative Commons 4, or similar, licensing. GPDC is a work in progress and is being continually expanded. At the time of publication, GPDC includes approximately 243 billion observations (or `data points') over 200,000 variables. The Knowledge Graph encoding contains approximately 3.7 trillion triples.

\subsection{Metadata and Coverage}
  Google Public Data Commons includes data not just about the external world, but also a suite of metadata inspection tools \cite{statvarexplorer}. It also includes data about its own coverage. \cite{country-coverage} provides the number of variables available for each country, at the country level. \cite{admin1-coverage} and \cite{admin2-coverage} provide the same information at the Administrative Area 1 level (U.S. equivalent of state) and Administrative Area 2 (U.S. equivalent of county). At a high level, excluding the U.S. and excluding countries with a population of less than 5 million, we have 9,500 variables at the country level, 600 variables at the Administrative Area 1 level and 705 variables at the Administrative Area 2 level. Table 2 shows this break down at a continent level. For the U.S., we have 146,000 variables at the country level, 121,000 variables at the state level and 98,500 variables at the county level. As table 2 and the charts in  \cite{country-coverage}, \cite{admin1-coverage} and \cite{admin2-coverage} show, we have deep coverage in the U.S., but much shallower coverage in countries in the Global South. We have significant efforts underway to improve our coverage in these places, but unfortunately, in many cases, the data is not available. 

In addition, Data Commons is dependent on the refresh rate and data accuracy provided by sources. For example, the U.S. census only releases its census count data once every 10 years, thus the most recent population and demographic data available comes from the 2020 release during the pandemic. As greatly covered in the news, the 2020 census dealt with many challenges ranging from operationally running the census during the pandemic to political. Nonetheless, we use much of the demographic data from the US Census in Data Commons. Other countries like Lebanon and India have either greatly delayed or completely stopped running a national census to avoid potential political problems. 

We do hope open source tools like Data Commons helps reduce the cost and skill barrier to sharing government data. Working closely with non-profits like TechSoup, we also hope that more non-governmental organizations and local organizations can help fill these data dark spots. 

\begin{table}[]
    \centering
    \begin{tabular}{|l|c|c|c|}
    \toprule
      Continent & Country & {Admin Area 1} & {Admin Area 2}     \\
      \midrule
      Africa   & 8898 & 15 & 410 \\
      Asia     & 8456 & 273 & 550 \\
      Europe   & 11810 & 2045 & 1505 \\
      North America & 9133 & 191 & 429 \\
      Oceania & 8195 & 77 & 477 \\
      South America & 9498 & 210 & 334 \\
      \bottomrule
    \end{tabular}
    \caption{Average number of variables at different geographical levels, per continent, excluding countries with a population of less than 5 million.}
    \label{tab:my_label}
\end{table}

\subsection{Other Data Commons}
\begin{figure}[H]
    \centering
    \begin{subfigure}{0.45\textwidth}
    \captionsetup{justification=centering}
        \centering
        \frame{
            \includegraphics[width=\linewidth]{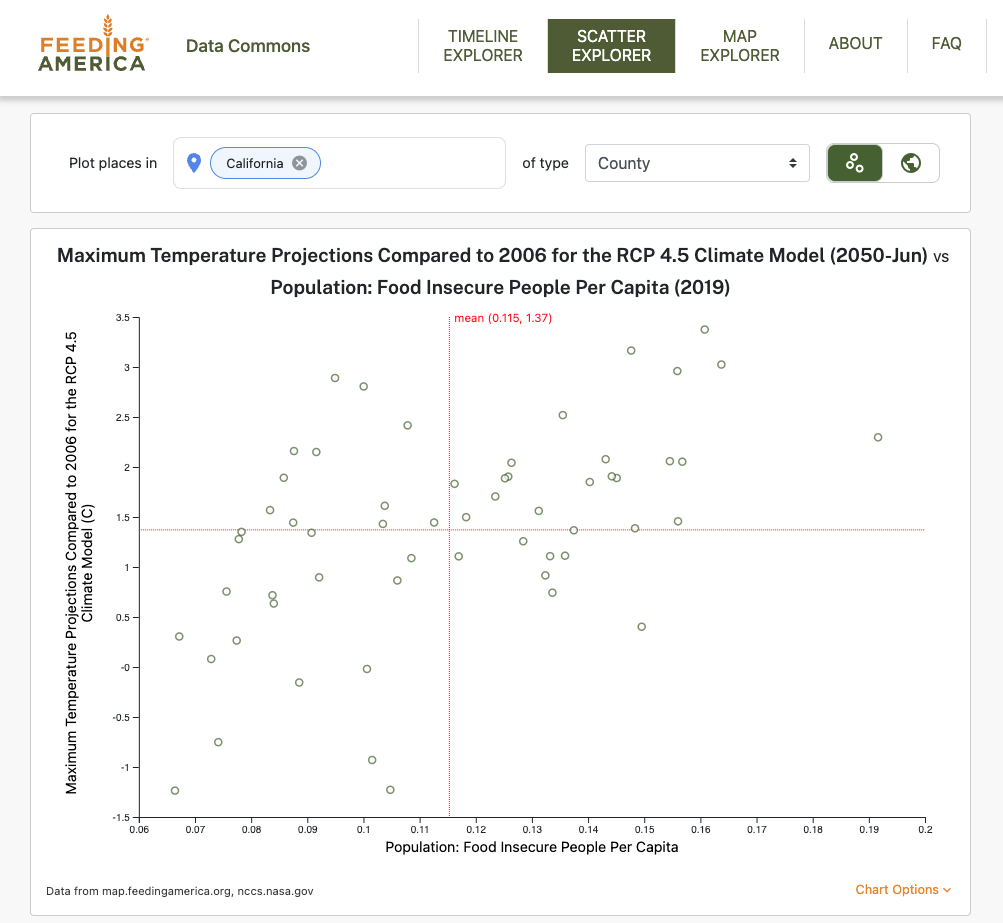}
        }
        \caption{}
        \label{fig:fa-dc}
    \end{subfigure}
    \hfill
    \begin{subfigure}{0.45\textwidth}
    \captionsetup{justification=centering}
        \centering
        \frame{
            \includegraphics[width=\textwidth]{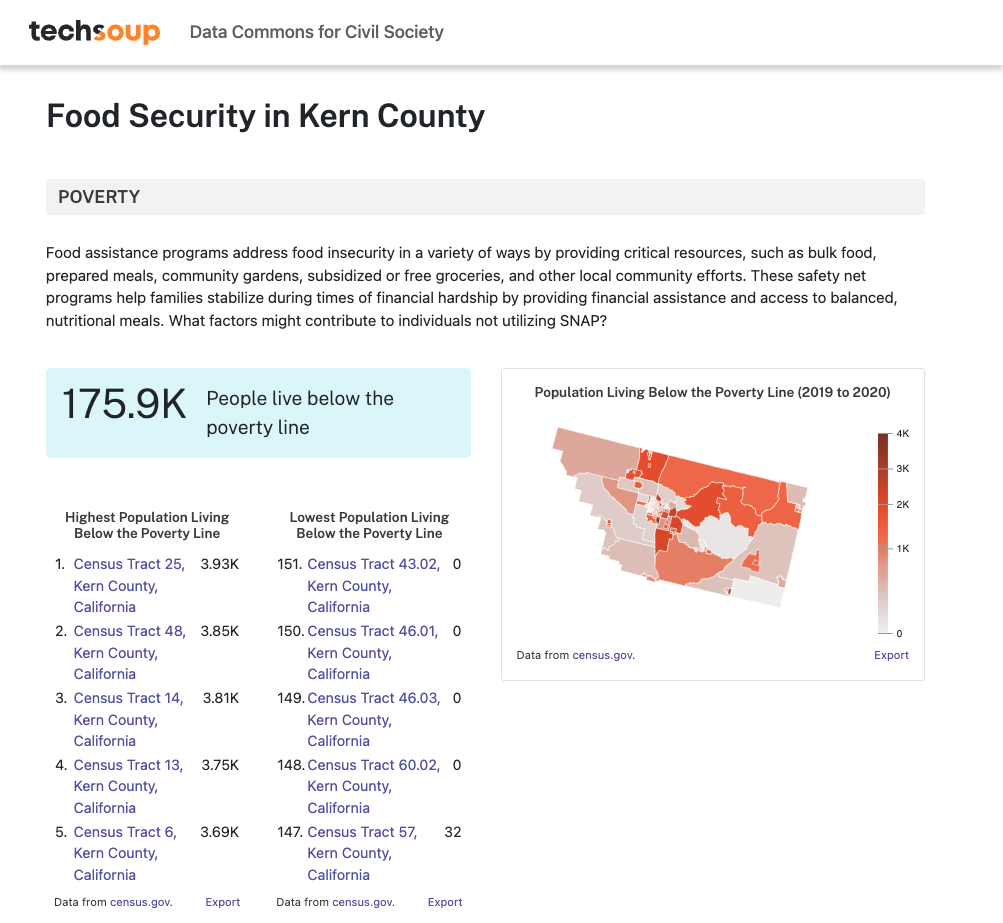}
        }
        \caption{}
        \label{fig:ts-dc}
    \end{subfigure}
    \caption{Applications built with Data Commons. \textbf{(a)} Feeding America meal gap data with Google Public Data Commons. \textbf{(b)} TechSoup food security overview for Kern County}
    \label{fig:dc-viz}
\end{figure}

\begin{itemize}
    \item  Feeding America: As mentioned earlier, Feeding America is a charitable organization that is deeply involved in the running of over fifty percent of the food kitchens in the US. Over the years, they have developed the Meal Gap index, a measure of food insecurity at the county level. They have set up a Data Commons at \cite{feedingamericadc} which contains their meal gap data. It is now possible to easily combine the meal gap data with data about various health conditions, demographics, and climate models. As seen in Fig \ref{fig:fa-dc}, by combining RCP4.5 temperature models from the IPCC with the meal gap index in counties in California, we can see that counties most at risk for climate change are already experiencing high rates of food insecurity. Digging a step deeper, we can quickly see that many of these counties are agricultural counties which may be the “canaries in the coal mine” for climate change. 
    
    \item TechSoup: TechSoup \cite{techsoupwebsite} is an organization that helps various charitable organizations with their technology needs. They have built an application that involves combining climate change data with data about health, jobs, poverty, etc. to determine which communities (at the census tract level) are most at risk. Further, the TechSoup Data Commons \cite{techsoupdc} combines this with data about local social service organizations to help determine the points from which these communities can be helped. See Fig \ref{fig:ts-dc} for an example of this in Kern County, California.
\end{itemize}

\section{Natural Language Interface}

  The recent rapid development of Large Language Models (LLMs) has enabled us to explore the use of natural language interfaces to Data Commons.  All of the examples listed above can now be answered by the natural language interface. Further, the conversational aspect of the interface is able to handle features such as context. An example of the kind of interaction that is now possible is given below: 
  \begin{itemize}
\item which countries in Africa have seen the most increase in electricity access
\item what is the increase in the GDP of these countries
\item how have the greenhouse gas emissions of these countries changed
\item how do the US and Germany compare to these countries
\end{itemize}

While there has been a lot of work in using LLMs to translate natural language queries into SQL, our application poses some challenges:
\begin{itemize}
    \item Typically, the schema for the structured data is specified in the prompt to the Chatbot. However, the extremely wide range of data in Google Public Data Commons makes this infeasible. While it is possible to conceptualize Google Public Data Commons as a relational table, such a table would have over 160,000 columns. 
    \item Many of the queries that users want to ask do not correspond directly to specific variables, but to broad topics that may be covered by a number of different variables. Consequently, the goal has to be to use natural language to help the user begin exploration of the space of data.
\end{itemize}

A forthcoming publication will cover our approach to these problems.

\section*{Acknowledgments}
We are very thankful to our colleagues at Google who have supported us over the years on this project. In particular, Urs H{\"o}elze and Sundar Pichai who helped initiate the project and Donald Harrison who provided a home for it. We have received great support from Kerry McHugh, Tessa Leuth, Luke Garske, Andrew Kim, Sheri Hunt and others in running the program. 
We also want to express our gratitude to many other people who have offered help for this project:

\subsection*{Team Members}
Srikanth Belwadi, Alex Chen, Tiffany Chen, Thejesh GN, Anush Kini, Dan Noble, Kiran Panesar, Pulkit Sachdeva, Keyur Shah, Sharad Shriram.

\subsection*{Interns}
Suhana Bedi, Antares Chen, Ian Costello, Christie Ellks, Padma Gundapaneni, Karen Huynh, Steve Lin, Lijuan Qian, Jay Shi, Senthamizhan V, Crystal Wang, Roger Wang, Hao Wu, Sinan Yumurtacı.

\subsection*{Contributing Colleagues}
Kate Brandt, Ari Gilder, Ben Gomes, Maggie Johnson, Emily Ma, Yael Maguire, Xiaofeng Mi, Sepi Moghadam, Ido Ohad, Jeffrey Oldham, Adriana Olmos, Mukund Raghavachari, Çınar Şahin, Ashwani Sharma, Anand Shukla, Corina Standiford, Alpana Ved, Balaji Venkatachari, Srinivasan Venkatachary, Shivakumar Venkataraman, Chealsea Wierbonski, Chuck Wu, Cong Yu.  

\subsection*{Advisors}
 Luiz André Barroso, Vint Cerf, Aneesh Chopra, Gary King, Tom Khalil, Arun Majumdar, Andrew Moore, Mark Musen, Alfred Spector, Hal Varian.

\subsection*{External Collaborators}
Hannah Druckenmiller,  Steve Lee, Mark Mollenkopf, Balasubramanian Narasimhan, Ram Rajgopal, Balaraman Ravindran, Aditi Sheshadri,  
Marnie Webb, Zach Whitman, Oliver Wise, Mike Yeaton.

\subsection*{Organizations}
Feeding America, Harvard University, Indian Institute of Technology Madras, Resources for the Future, Stanford University, TechSoup, United Nations Department of Statistics

\bibliography{library}

\bibliographystyle{IEEEtran}

\newpage
\begin{appendices}
\section{Google Data Commons Sources}

  To jumpstart the Ecosystem, we have built the Google Public Data Commons of publicly available data from over 200 different sources, covering thousands of data sets. Some of these include

\begin{itemize}
\item Demographics: The demographics data spans over 35.5K statistical variables. We collect our demographics information from sources such as: Australian Bureau of Statistics, Brazilian Institute of Geography and Statistics (IBGE), Census of India, Colombia DANE National Administrative Department of Statistics, DataMeet, EU Eurostat, Google, India National Informatics Centre, National Institute of Statistics and Censuses (INDEC), Open Data for Africa, Opportunity Insights, Organization for Economic Co-operation and Development (OECD), Portal Site of Official Statistics of Japan (e-Stat), Statistics Canada, U.S. Bureau of Transportation Statistics, U.S. Census Bureau, U.S. Center for Disease Control and Prevention (CDC), U.S. Department of Housing and Urban Development (HUD), U.S. Federal Election Commission (FEC), U.S. National Center for Education Statistics, United Nations (UN), Wikimedia Foundation, and World Bank.
\item Economics: The economics data spans over 67.5K statistical variables. We collect our economics information from sources such as: Australian Bureau of Statistics, EU Eurostat, Food and Agriculture Organization of the United States, India Ministry of Statistics and Programme Implementation, Organization for Economic Co-operation and Development (OECD), Portal Site of Official Statistics of Japan (e-Stat), Reserve Bank of India, U.S. Bureau of Economic Analysis (BEA), U.S. Bureau of Labor Statistics (BLS), U.S. Census Bureau, U.S. Department of Housing and Urban Development (HUD), U.S. Department of Labor (DOL), U.S. Federal Reserve, and World Bank.
\item Education, housing, etc.: The education, housing, and commute data spans over 42.4K statistical variables. We collect our education, housing, and commute information from sources such as: Brazilian Institute of Geography and Statistics (IBGE), California Assessment of Student Performance and Progress, Colombia DANE National Administrative Department of Statistics, EU Eurostat, India Ministry of Education, Open Data for Africa, Portal Site of Official Statistics of Japan (e-Stat), U.S. Census Bureau, U.S. Department of Education (ED), and U.S. National Center for Education Statistics.
\item Public Health: The health data spans over 20.5K statistical variables. We collect our health information from sources such as: Australian Bureau of Statistics, City Controller, City of Los Angeles, Colombia DANE National Administrative Department of Statistics, Dartmouth Atlas Project, DataMeet, EU Eurostat, Google, India National Health Mission, India National Sample Survey, National Institute of Statistics and Censuses, Open Data for Africa, Our World in Data, Portal Site of Official Statistics of Japan (e-Stat), The Central Bureau of Statistics, Indonesia, The New York Times, U.S. Census Bureau, U.S. Center for Disease Control and Prevention (CDC), U.S. Drug Enforcement Agency (DEA), World Bank, and World Health Organization (WHO).
\item Climate, Sustainability: The sustainability data spans over 12.7K statistical variables. We collect our sustainability information from sources such as: Climate Trace, Data Commons, Dynamic World Project, European Union (EU) Copernicus, Federal Emergency Management Agency (FEMA), Global Land Ice Measurements from Space (GLIMS), India Central Pollution Control Board, India Energy Dashboard, India Water Resources Information System, Resources for the Future (RFF), Stanford University, U.S. Bureau of Transportation Statistics, U.S. Center for Disease Control and Prevention (CDC), U.S. Energy Information Administration (EIA), U.S. Environmental Protection Agency (EPA), U.S. National Aeronautics and Space Administration (NASA), U.S. National Oceanic and Atmospheric Administration (NOAA), U.S. National Renewable Energy Laboratory (NREL), U.S. National Wildlife Coordinating Group, United Nations (UN), United States Geological Service (USGS), Wildland Fire Interagency Geospatial Services, and World Bank.
\item Biomedicine:  Biomedical data covers a wide range, from data about public health, diseases, drugs to sequences and proteins. Unfortunately, this data is in a number of silos, in different schemas and formats. Consequently, though there is an overlap in the entities that these different databases refer to, it is very difficult to combine the data across them. Biomedical Data Commons is an effort to create a single Knowledge Graph out of all this data. Currently the Biomedical Data Commons includes data from Encode, ClinVar, UniProd, HUPO, HMDB, ICD-10 and others.
\end{itemize}

\end{appendices}

\end{document}